\documentclass[twocolumn, traditabstract]{aa}
\usepackage{amsmath}
\usepackage{graphicx}   
\usepackage{bm}         
\usepackage{natbib}
\usepackage{etoolbox}
\usepackage{rotating}
\usepackage{array}
\usepackage{booktabs}
\usepackage{amssymb}
\usepackage{gensymb}
\usepackage{hyperref}
\usepackage{wrapfig}
\usepackage{ulem}
\usepackage{booktabs}
\usepackage{etoolbox}
\usepackage{lipsum}
\usepackage{enumitem}
\usepackage[dvipsnames]{xcolor}
\usepackage{caption}
\usepackage{linenoaa}
\usepackage{etoolbox}

\AtBeginEnvironment{thebibliography}{%
    \let\linenumbers\relax
    \nolinenumbers
}

\definecolor{posgreen}{HTML}{008000}
\setlist[enumerate]{itemsep=0mm}
\usepackage[table,x11names]{xcolor}

\begin{document}

\title{Results from a Stellar Occultation by (623) Chimaera}
\author{
Phillip Oakey\inst{1,2}
\and
Keya Garg\inst{3}
\and
Andrey Moore\inst{2}
\and
Altony Foote\inst{2}
\and
Michael F. Skrutskie\inst{2}
\and
Anne J. Verbiscer\inst{2}
\and
Matthew Nelson\inst{2}
\and
Marc W. Buie\inst{4}
\and
Diego Gomez\inst{2}
\and
Lauren Daszynski\inst{2}
\and
Charlie Dorsey\inst{2}
\and
Roger Venable\inst{5}
\and
Vadim Nikitin\inst{5}
\and
Joan Dunham\inst{5}
\and
David Dunham\inst{5}
\and
Nadara Hudson\inst{2}
\and
Colin Surovell\inst{2}
\and
Norman Carlson\inst{5}
\and
Georgia Clickner\inst{2}
\and
James Beck\inst{3}
\and
Mark Chernov\inst{2}
\and
Tristan Howdyshell\inst{2}
\and
Paul D. Maley\inst{5,6}
\and
Ted Barnes\inst{7}
\and
John DiPasquale\inst{3}
\and
Casey York\inst{2}
\and
Ted Blank\inst{5}
\and
Henk de Groot\inst{8}
}
\institute{
Department of Astronomy and Physics, Uppsala University, Uppsala, Uppsala Län, 751 05, Sweden
\and
Department of Astronomy, University of Virginia, Charlottesville, VA, 22904, USA
\and
School of Engineering and Applied Science, University of Virginia, Charlottesville, VA, 22904, USA
\and
Planetary Science Institute, Tucson, AZ, 85719, USA
\and
International Occultation Timing Association, P.O. Box 20313, Fountain Hills, AZ, 85269-0313, USA
\and
NASA Johnson Space Center Astronomical Society, Carefree, AZ, 85377, USA
\and
Tamke-Allan Observatory, Rockwood, Tennessee, 37854, USA
\and
Dutch Occultation Association, Zonnebloemhof 3, 7732 BL Ommen, The Netherlands
}

\abstract{
We present results from a stellar occultation by the asteroid (623) Chimaera on October 21st, 2025, when it passed in front of V magnitude 13.86 star UCAC4 626-004193. Chimaera is the largest asteroid in the Chimaerid asteroid family and will be subject to a spacecraft flyby in the 2030s. Eighteen telescopes were deployed, spanning the United States from Arizona to North Carolina, incrementally spaced by $\sim4.4$\,km with respect to the cross-track of the shadow. In total, sixteen telescopes collected useful data. This multi-chord observation, when compiled into a single sky-plane plot, suggests a profile which is well-described by an elongated elliptical fit, with a semi-major axis $a=26.2\pm0.8$ km and a semi-minor axis $b=16.0\pm1.0$ km. The projected area from ellipse fitting is 1316\;km$^2$, and a limb fit is calculated to enclose 1312 km$^2$. Photometry collected in the \textit{Gaia} G band over a seven-month period coinciding with the occultation corresponds to a geometric albedo $p_G=0.059\pm0.005$; comparing with archival occultation observations and direct observations gives $p_G=0.051\pm0.004$ and $p_V=0.051\pm0.004$. We compare occultation sky-plane projections with 3-D modeling from inversion techniques and demonstrate that the models can accurately predict the orientation of Chimaera at the time of each occultation.
}


\maketitle
\nolinenumbers

\section{Introduction}
\label{sec:introduction}

Stellar occultations, which occur when Solar System bodies transit stars, have been shown repeatedly to be one of the most powerful tools to probe minor planets from the surface of the Earth \citep{1961MNSSA..20..134E, 1987Icar...72..507M, 2023Natur.614..239M}. Precise timing (millisecond or sub-millisecond) of the ingress and egress of the occulting body's shadow provides the most accurate astrometric measurements obtainable through atmosphere, and when combined with multiple telescopes distributed throughout the cross-track of the shadow, can be used to probe topography on the kilometer scale. Occultations have also been used to detect satellites around asteroids \citep{2022DPS....5451203B, 2022MPBu...49....3G}

The inner main-belt asteroid (623) Chimaera (hereafter Chimaera) orbits the Sun in an orbit $\sim2.47$ au in radius, with an eccentricity of 0.112 and a relatively high inclination of $14.1\degree$ (gathered from the JPL$\#$108 ephemeris, \citealt{1997DPS....29.2106C}). It occupies an orbit very close to the 3:1 resonance with Jupiter at 2.5 au (known as the 3:1 Kirkwood gap). This specific mean motion resonance in the main asteroid belt, along with the \textit{v}$_6$ secular resonance, is hypothesized to be the progenitor of the majority of inner-Solar System asteroidal material, including meteors and meteorites \citep{1990Icar...87...78Y, 1995Icar..115...60M, 1995Icar..114...33M}. 

\begin{table*}[htbp]
    \centering
    \footnotesize
    \begin{tabular*}{\textwidth}{c|c|c|c|c|c|c|c|c|c}
        \hline
        \hline
        Number & Name     &  Diameter       & Absolute mag. & Rotation period & Geometric albedo        & \textit{a} & Axial tilt & e     & \textit{i} \\
               &          & [km]            & [V band]      & [h]             & [V band]                & [au]       & [degree]   &       & [$\degree$] \\
        \hline
        623    & Chimaera & $44.1\pm1.0$    & 10.83         & 14.6241         & $0.0372\pm0.002$        & 2.461      & 78         & 0.112 & 14.1                   \\
        \hline
        \hline
    \end{tabular*}
    \caption{Physical characteristics of (623) Chimaera from the literature preceding this manuscript. Diameter, absolute magnitude, and geometric albedo are gathered from a series of eight (IRAS) observations, rotation period is gathered from shape modelling from inversion techniques on \textit{Gaia} data from 2014-2017 \citep{2023A&A...675A..24D}.}
    \label{tab:Chimaera_params}
\end{table*}

Chimaera, for many reasons, is a unique asteroid. At $\sim78\degree$, the spin axis obliquity of Chimaera \citep{2024AGUFMP23G...07S} is comparable to that of the giant planet Uranus ($82\degree$). The asteroid has become of particular interest when it was identified as the parent of the high-inclination, eponymous Chimaerid asteroid family \citep{2015PDSS..234.....N}. This family is composed of 108 identified member asteroids which are predominantly the primitive C-type in the taxonomical scheme of \citet{1984PhDT.........3T}, as suggested through a survey of twenty Chimaerids with the OSIRIS spectrometer on the 10.4 m Gran Telescopio Canarias \citep{2019A&A...630A.141M}. This spectral analysis demonstrates that the Chimaerid family possesses a non-negligible fraction of hydrated bodies, indicating that at some point in the history of the family, low temperature (sub-320 K) liquid water was involved in their collective chemical evolution, generating carbonates, hydroxides, oxides, phyllosilicates, sulfates, and other compounds \citep{2014acm..conf..165F}. The existence of this large family of asteroids related to the parent Chimaera indicates a collisional event at some point in Chimaera's history, in which large pieces of debris (the largest, (2642) Vesale, with a diameter of 7.4 km) were ejected into different heliocentric orbits. It is possible that the abnormal spin axis obliquity of Chimaera is a result of this prehistoric collision, as has long been suggested for Uranus \citep{1966SvA.....9..987S, 2018ApJ...861...52K}. Interestingly, it has been suggested that there is a low chance that the asteroid (162173) Ryugu, visited by the Hayabusa 2 mission in 2018, may be a member of the Chimaera family \citep{2019A&A...630A.141M}.

Chimaera is of further interest due to its selection as a target for the upcoming \textit{Emirates Mission to the Asteroid Belt} (EMA) spacecraft, currently under construction by the United Arab Emirates Space Agency (UAESA) and expected to launch in 2028. This spacecraft will fly by six asteroids before entering orbit around the seventh, (269) Justitia, before deploying a probe which will perform an uncontrolled descent to the surface of the asteroid \citep{2024cosp...45..261A, 2026arXiv261026581B}. Chimaera is the second largest of the asteroids to be surveyed by this spacecraft. Successful stellar occultation measurements have been conducted for two of the EMA target asteroids, (269) Justitia \citep{2025PSJ.....6...61B} and (59980) Moza (Oakey et al. in prep.), probing their size, scale, and the possibilities for small satellites around these bodies. 

General information for Chimaera can be found in Table \ref{tab:Chimaera_params}.  The geometric albedo $p_V=0.0372\pm0.002$ presented here is derived from Infrared Astronomical Satellite (IRAS, \citealt{2004PDSS...12.....T}) observations, which is calculated from an estimate of the infrared radiometric albedo; this diameter estimate possesses unconstrained systematic errors on the order of $10-20\%$ \citep{2002aste.book..205H}, and therefore better constraints on asteroid dimensions are desired. 

In this work, we will present: the methodology used to predict an occultation by (623) Chimaera (Section \ref{sec:predictions}); the organization of efforts for an observation campaign, a description of the telescopes and detectors utilized for the measurement at each station, along with the data from the multi-chord observation of the occultation (Section \ref{sec:observations}); a discussion of the morphology, size, and shape of Chimaera from this observation and prior occultation results (Section \ref{sec:discussion}). All dates presented in this work are in the format \textit{YYYYMMDD} (i.e. July 12, 1947 would be presented as 19470712).

\section{Predictions}
\label{sec:predictions}

The advent of \textit{Gaia} (Global Astrometric Interferometer for Astrophysics) measurements for the astrometric positions and proper motions of stars \citep{2021A&A...649A...1G} has enormously improved the precision of stellar occultation predictions. To generate predictions for occultations by Chimaera, we used stellar astrometry from \textit{Gaia} DR3 (down to V mag 16) and the JPL\#108 ephemeris for Chimaera from the Horizons database \citep{1997DPS....29.2106C}, which is compiled from a 118 year observation period. These predictions made use of the \texttt{Occult4} software, which also maintains a database of all occultation observations recorded by the International Occultation Timing Association (IOTA) \citep{2020MNRAS.499.4570H}. This search yielded one ideal occultation within the desired time period ($\sim2025-2026$) and the desired geographical constraints for a large deployment of telescopes from the University of Virginia in Charlottesville, Virginia, USA. On 20251021, Chimaera would pass in front of UCAC4 626-004193, a V magnitude 13.86 star in the constellation Andromeda. The shadow was predicted to begin its transit of the Earth over the Atlantic Ocean, crossing the southeastern, south-central, and southwestern United States before entering the Pacific Ocean, crossing over a portion of the island of Hawai`i prior to termination in the southern Pacific. The parameters for this specific occultation can be found in Table \ref{tab:occ_params}.

\begin{table*}[htbp]
    \centering
    \footnotesize
    \begin{tabular*}{\textwidth}{c|c|c|c|c|c|c|c}
        \hline
        \hline
        Date     & Star       & Stellar V mag. & Chimaera V mag. & R.A.            & Dec.            & Duration & Shadow speed  \\ 
                 & (UCAC4)    &               &                & [hh:mm:ss.ssss] & [dd:mm:ss.sss]  & [s]      & [km s$^{-1}$] \\
        \hline
        20251021 & 626-004193 & 13.86         & 13.8           & 01:21:50.9813   & $+$35:17:08.862 & 4.7      & 9.28          \\
        \hline
        \hline
    \end{tabular*}
    \caption{Parameters for the occultation by Chimaera presented in this work, as calculated with \textit{Gaia} DR3 astrometry for stars to V magnitude 16 and the JPL\#108 ephemeris for Chimaera. The ``Duration'' column provides the predicted maximum duration of the occultation using estimates for the asteroid diameter from IRAS. The ``Shadow speed'' is the calculated velocity of the shadow from a geocentric point of view.}
    \label{tab:occ_params}
\end{table*}

\section{Observations}
\label{sec:observations}

\subsection{Campaign on 20251021}
\label{sec:campaign}


\begin{figure*}
    \centering
    \includegraphics[width=\textwidth]{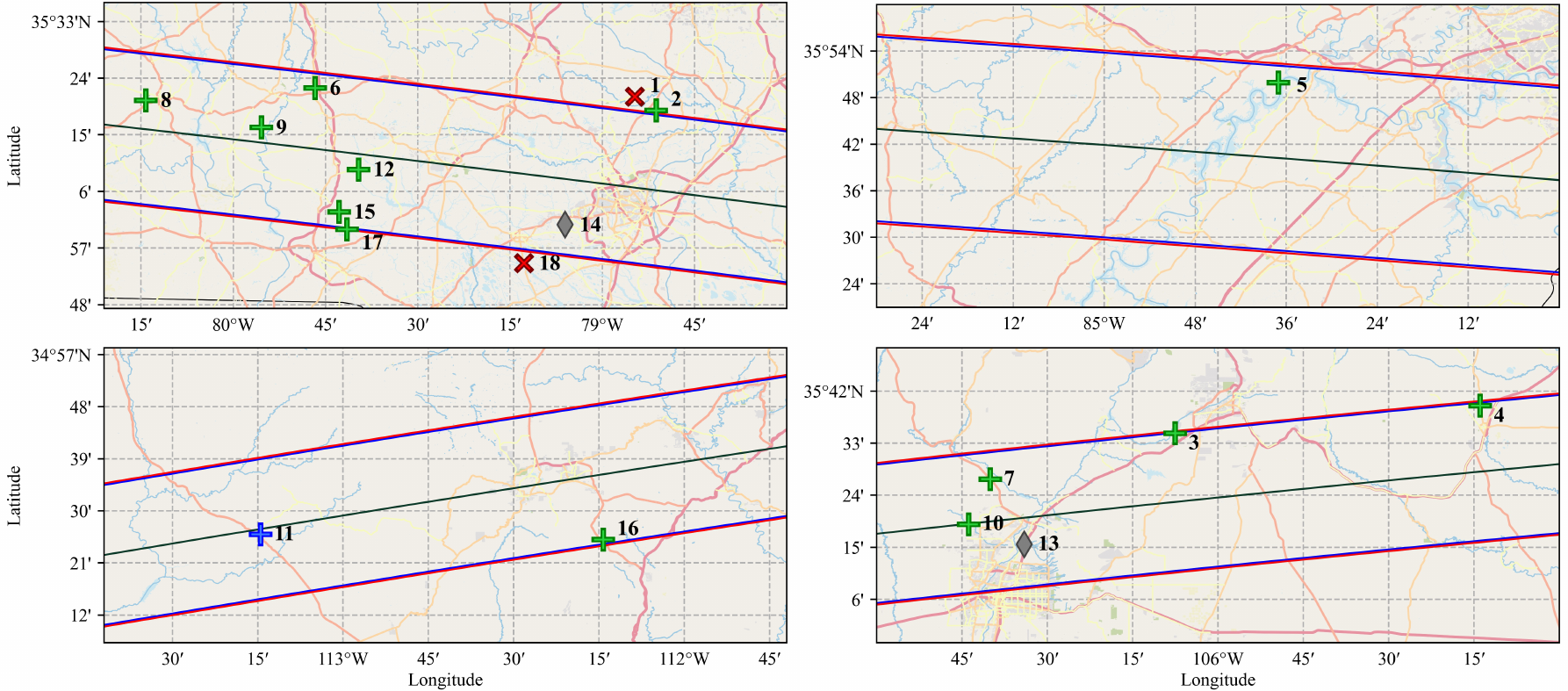}
    \caption{Locations of observing sites across the continental United States. From top left, clockwise, the sites are located in the U.S. states of North Carolina, Tennessee, New Mexico, and Arizona. The dark green line in each panel shows the predicted center path from orbit JPL\#108, the blue lines show the predicted shadow edge presuming the IRAS-derived diameter of $44.09\pm1.0$ km, and the red lines are the 1-$\sigma$ region of uncertainty for the ephemeris, not accounting for geometric uncertainty. Crosses denote stations where an occultation was detected (with a blue cross for a half-detection; see §\ref{sec:campaign}), ``x''s denote stations which did not detect an occultation, and diamonds denote stations where data were not collected.}
    \label{fig:sites}
\end{figure*}

The nature of this occultation meant that measurements were primarily supported by volunteer observers with a variety of telescopes, detectors, and Global Positioning System (GPS) timing devices. Some telescopes lacked motorized clock drives, and observed the occultation by pointing to the altitude and azimuth of the field at event time. The primary cameras used in this campaign are as follows: the QHY174M-GPS, which uses a Sony IMX174 detector (IMX174 in tabular format; Tab. \ref{tab:20251021_telescopes}), a scientific complementary metal-oxide-semiconductor (sCMOS) camera which directly embeds GPS time from a receiver into the .FITS file headers; the Watec 910HX (WAT910 in tabular), an analog camera with timestamps inserted into video feed via the IOTA Video Time Inserter device; the ZWO ASI432MM (uncooled) or Apollo-M MAX Pro (cooled), which both utilize a Sony IMX432 detector (IMX432 in tabular) which are sCMOS cameras without direct timing metadata; the similar ZWO ASI178MM (IMX178 in tabular, uncooled, also without direct timing metadata),  and the ASTRID sCMOS camera with a Sony IMX296 detector (IMX296 in tabular), designed specifically for occultation science and obtains GPS metadata from a ``shield'' built into the larger ASTRID system. The QHY174M-GPS, ZWO ASI432/178MM, and ASTRID sCMOS cameras have near zero dead-time. Most of the detectors were operated with the \texttt{SharpCap} software running in 16-bit mode, with varying settings for gain and offset (gain between 200-300 across all systems, offset generally set to 50); with the exception of the ASTRID systems which were operated through their proprietary software on a Raspberry Pi processor, with a gain of 16. The ZWO ASI432/178MM and Apollo-M MAX Pro cameras without intrinsic time control require a separate method to discipline the computer's system time. This is achieved through a network time protocol software, receiving GPS signal through a pulse-per-second GPS module, calibrated via the established ``flash-tag'' technique \citep{2020JOA....10a...9G}.  The timing of the Watec 910HX camera is corrected for a systematic offset between the beginning of the exposure and the application of the GPS timestamp. Table \ref{tab:20251021_telescopes} lists the detector used for each telescope during the 20251021 campaign.

\begin{figure*}
    \centering
    \includegraphics[width=\textwidth]{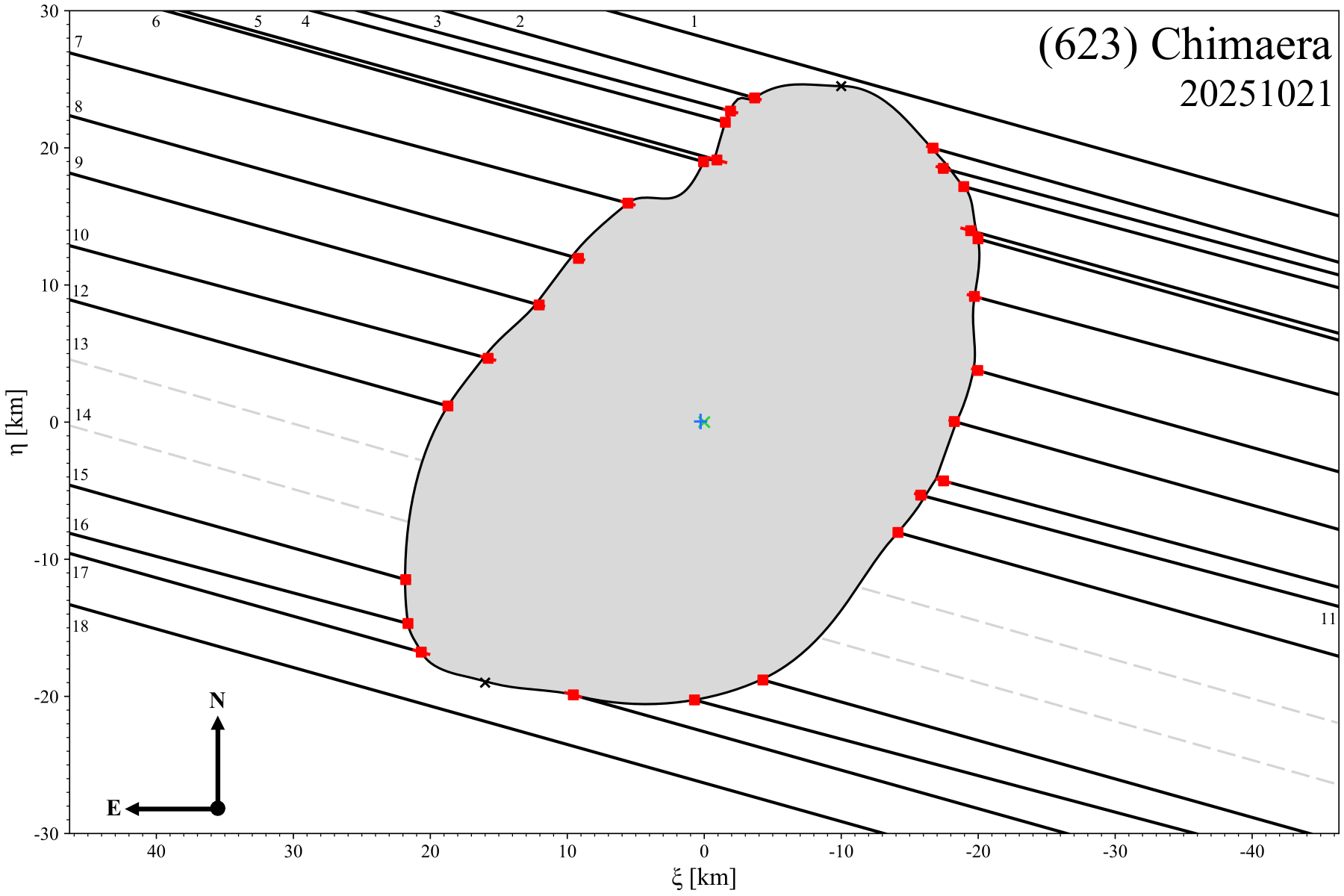}
    \caption{Sky-plane plot for (623) Chimaera using data collected during the 20251021 event. Stellar visibility marked in black with uncertainty in dis- and re-appearance marked in red. The green ``x'' in the center denotes the predicted center from the JPL\#108 ephemeris, and the blue cross marks the center of ellipse of best-fit. The dashed gray lines denote chords which did not record data. Control points to maintain a sensible limb are plotted as black ``x''s.}
    \label{fig:skyplaneplot}
\end{figure*}

The Occultation Group at the University of Virginia (hereafter the Occultation Group) prepared an array of eight telescopes, ranging in size from 25.4 cm to 35.6 cm and with varying detectors and timing systems (see Tab. \ref{tab:20251021_telescopes} for details). The Occultation Group submitted a call for observations to IOTA in advance of this event and several volunteer observers indicated their interest in participating in a coordinated campaign. This allowed for observing sites to be selected in a manner which evenly spaced the one-dimensional cross-track telescope distribution to deliver consistent resolution across the body (the intersection of a given telescope with the cross-track of the shadow is known commonly as a ``chord'', and a given chord will follow the two-dimensional projection of the shadow across the surface of Earth. Any telescope which observes the occultation from any point along a desired chord will observe the same topography of the asteroid as any other telescope on the same chord, neglecting the rotational effects of the asteroid. See chords 5 and 6 in Figs. \ref{fig:sites} and \ref{fig:skyplaneplot} for an example.). After an initial waiting period, the total number of telescopes planning to observe this event rose to fifteen. We proposed spacing thirteen of these fifteen chords evenly, centered on the predicted ephemeris from JPL$\#$108, to achieve a spatial resolution of 4.4 km. The remaining two telescopes (chords 4 and 14 in Fig. \ref{fig:skyplaneplot}) were tactically placed $\sim$ 1-2 kilometers within the interior of the predicted shadow edge, in order to more sensitively probe the sharp topography near the limb of the shadow. Additional observers were advised to observe wherever possible, but with a preference to place themselves near the northern or southern limbs. There would be three additional telescopes which joined in this manner (chord IDs 2, 5, and 11, as seen in Table \ref{tab:20251021_telescopes} and Figure \ref{fig:skyplaneplot}). The information gained from the proximity of these ``volunteer'' telescopes to planned chords is particularly interesting and is addressed in Section \ref{sec:morphology}.

Two chords, IDs 13 and 14 (in New Mexico and North Carolina, respectively), experienced issues with their chosen sites and were unable to collect data as anticipated. This leads to an unfortunate situation in which there is a large gap, approximately 13 km in width, in the coverage of the asteroid on the southern portion of the profile (marked by dashed gray lines in Fig. \ref{fig:skyplaneplot}). The field laptop used for chord ID 11 ran out of storage on the disk at 07:20:58.4907 (frame number 1706), which regrettably occurred during the occultation, meaning there is no measurement for stellar reappearance from this chord. This can be seen graphically in Figure \ref{fig:lightcurves} and software limitations mean that this chord must be excluded from consideration for the best-fit ellipse; see Section \ref{sec:ellipse}. 

Chord 11 used a Watec 910HX detector with the well-tested IOTA Visual Time Inserter (VTI), which is routinely used for occultation science and which was also employed for chords 14 and 18 in this work. There is no reason to suspect the timing from this chord was compromised, as the ingress is in good agreement with Chords 10 and 12.

Observations were reduced with the \texttt{PyMovie} aperture photometry package \citep{2019JOA.....9d...9A}. Timing information was extracted using the associated \texttt{PyOTE} package. Figure \ref{fig:lightcurves} displays the normalized lightcurves for each observation involved in the 20251021 campaign, centered on the predicted occultation mid-time for each observer's longitude and latitude. Chord 9 witnessed a cloud pass over the field prior to the event and the lightcurve has been trimmed accordingly.

\begin{figure*}
    \includegraphics[width=\textwidth]{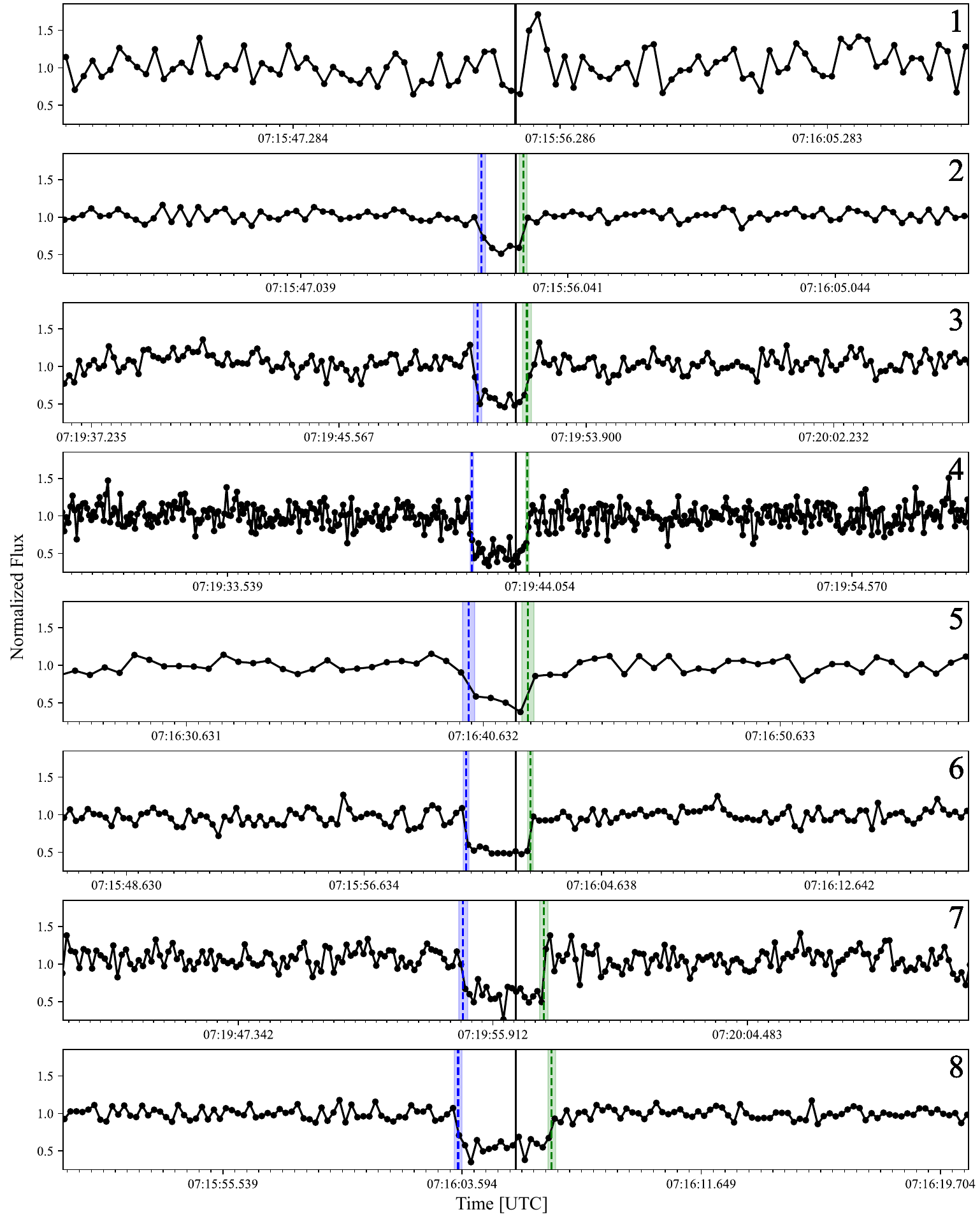}
    \caption{Normalized lightcurves for each chord observed for the occultation by (623) Chimaera on 20251021. Ingress of Chimaera is marked in blue and egress is marked in green with associated 3-$\sigma$ uncertainty for visibility. The predicted mid-time for the latitude and longitude of the observation is shown as a vertical black line.}
    \label{fig:lightcurves}
\end{figure*}

\addtocounter{figure}{-1}
\begin{figure*}
    \ContinuedFloat
    \includegraphics[width=\textwidth]{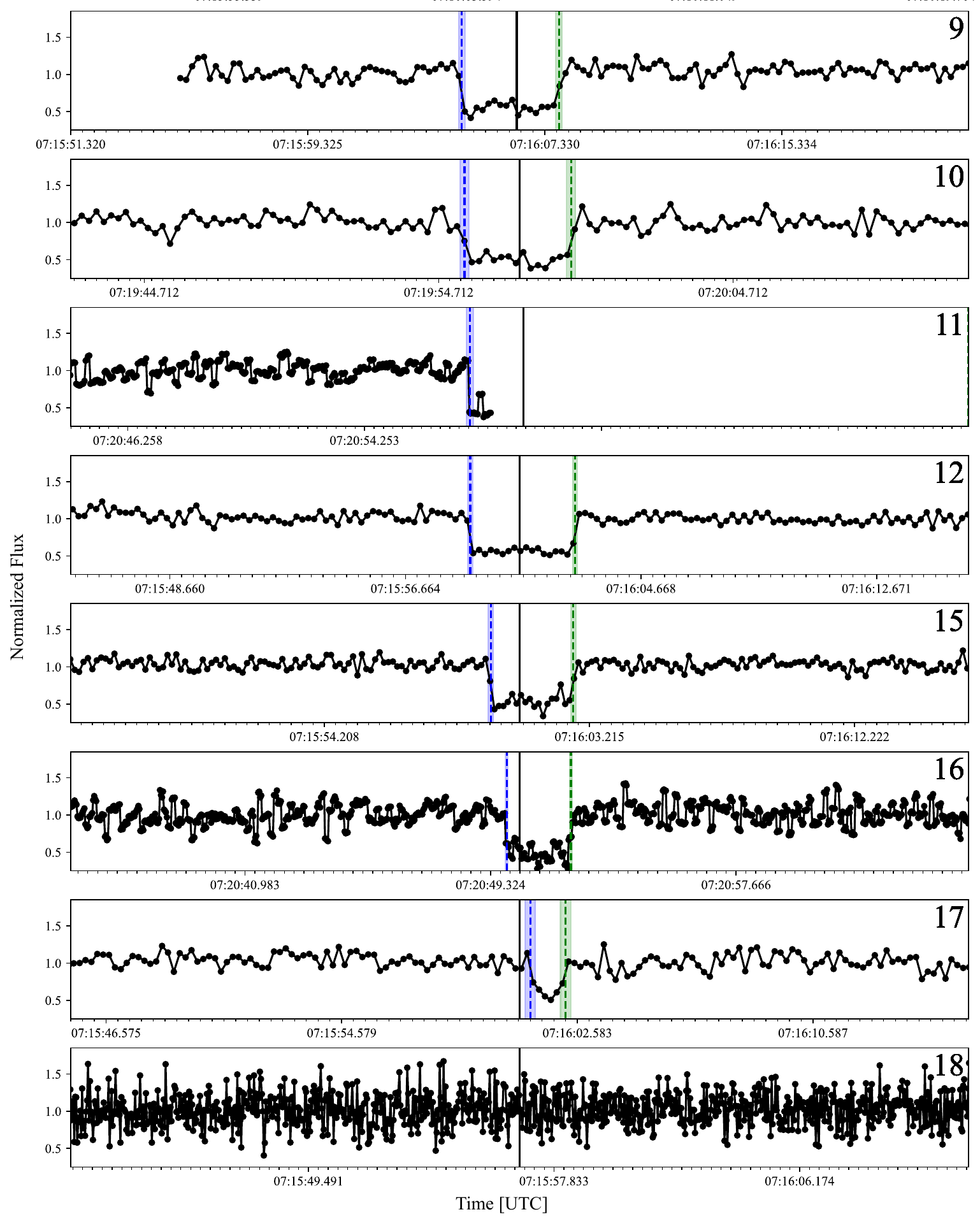}
    \caption{\textbf{(continued)} Transient clouds caused gaps in the data for chord 9; chord 11 experienced disk failure during recording.}
\end{figure*}

\begin{table*}[htbp]
    \centering
    \scriptsize
    \begin{tabular*}{0.97\textwidth}{c|c|c|c|c|c|c|c|c}
        \hline
        \hline
        ID & Aperture      & Detector      & Observer(s)                            & Latitude      & Longitude     & Altitude & Result                         & Notes  \\
           & $\text{[cm]}$ &               &                                        & [dd.dddddd]   & [ddd.dddddd]  & [m]      &                                &        \\
        \hline
        1  & 35.6 & IMX178 & K. Garg, D. Gomez, P. Oakey & 35.351222 & -078.911917 & 67   & \textcolor{red}{NEG} & \\
        2  & 25.4 & IMX432 & P. Oakey & 35.314500 & -078.854222 & 56 & \textcolor{posgreen}{POS} & \\
        3  & 23.5 & IMX296 & N. Carlson & 35.578972 & -106.125944 & 1885 & \textcolor{posgreen}{POS} & \\
        4  & 27.9 & IMX174 & V. Nikitin & 35.657972 & -105.232278 & 2019 & \textcolor{posgreen}{POS} & \\
        5  & 20.0 & IMX174 & T. Barnes & 35.832417 & -084.617861 & 324 & \textcolor{posgreen}{POS} & \\
        6  & 27.9 & IMX174 & M. Chernov, T. Howdyshell, C. Surovell & 35.374028 & -079.778333 & 188 & \textcolor{posgreen}{POS} & \\
        7  & 25.4 & IMX296 & J. Dunham, D. Dunham & 35.446370 & -106.666790 & 1632 & \textcolor{posgreen}{POS} & \\
        8  & 27.9 & IMX174 & J. DiPasquale, C. York, J. Beck & 35.341583 & -080.237722 & 160 & \textcolor{posgreen}{POS} & \\
        9  & 27.9 & IMX432 & A. Moore, A. Salinas, K. Elavarasu & 35.269611 & -079.924417 & 146 & \textcolor{posgreen}{POS} & \\
        10 & 20.0 & IMX296 & T. Heisey & 35.316167 & -106.731056 & 1752 & \textcolor{posgreen}{POS} & \\
        11 & 20.0 & WAT910 & P. Maley & 34.432306 & -113.242110 & 778 & \textcolor{blue}{POS-D} & Ingress only \\
        12 & 35.6 & IMX432 & N. Hudson, C. Dorsey, G. Clickner & 35.158333 & -079.660833 & 160 & \textcolor{posgreen}{POS} & \\
        13 & 20.3 & IMX296 & D. Dunham, J. Dunham & 35.258373 & -106.567415 & 1543 & DNO & Site issues \\
        14 & 35.6 & WAT910 & R. Venable & 35.011692 & -079.145680 & 79 & DNO & Site issues \\
        15 & 30.5 & IMX432 & M. Skrutskie & 35.045806 & -079.713639 & 110 & \textcolor{posgreen}{POS} & \\
        16 & 25.0 & WAT910 & T. Blank & 34.417250 & -112.247750 & 1387 & \textcolor{posgreen}{POS} & \\
        17 & 27.9 & IMX432 & A. Foote, L. Daszynski, S. Welch & 34.999722 & -079.692222 & 106 & \textcolor{posgreen}{POS} & \\
        18 & 35.6 & WAT910 & R. Venable & 34.910306 & -079.212417 & 68 & \textcolor{red}{NEG} & \\
        \hline
        \hline
    \end{tabular*}
    \caption{List of observers and equipment with geographical distribution for the 20251021 campaign to observe (623) Chimaera. Disappearance and reappearance times for each chord can be found in Appendix \ref{app:d&r}.}
    \label{tab:20251021_telescopes}
\end{table*}

\subsection{Archival Occultation Data}
\label{sec:prior_occs}

There is a valuable history of occultations by (623) Chimaera which have been stored in public archives kept within the \texttt{Occult4} framework. We refer to these observations to help contextualize the results obtained on 20251021. In total, five prior occultations by Chimaera have been observed over a sixteen year epoch, spanning from 2008 to 2024. Of these five observations, three were measured by single chords, one was measured by three chords, and one was measured by four. We present the relevant information for these occultations and their resulting sky-plane plots in Table \ref{tab:prior_occ_params} and Figure \ref{fig:ellipses}, respectively.

\subsection{3-D Modelling}
\label{sec:3dmodel}

There has been extensive work in recent decades to produce a catalog of three-dimensional shape models for the population of minor planets in the Solar System. The Database of Asteroid Models from Inversion Techniques (hereafter DAMIT) constructs shape models for all possible asteroids using photometric datapoints from \textit{Gaia} collected from 20140914 to 20170521 \citep{2010A&A...513A..46D}. The model and associated parameters can prove very useful for interpreting the results from stellar occultations, particularly for occultations in which only a single positive chord was recorded. The techniques used to generate the DAMIT model for Chimaera also produce a period for the asteroid and, for a given occultation mid-time, a predicted rotation phase curve, which is invaluable for contextualizing the predicted orientation of the asteroid. The period reported by DAMIT (14.6241 h) is not necessarily in agreement with the historical record in the Small Body Database \citep[14.635 h]{2009Icar..202..134W}, but is in good agreement with the value derived from independent photometry discussed in Section \ref{sec:photometry} (14.626646 h). We use the predicted orientation, shape, and size from DAMIT modelling to provide initial guesses for our ellipse fitting in Section\,\ref{sec:ellipse}.

Notably, the authors of the DAMIT remark that the nature of lightcurve inversion means it is very difficult to reveal non-convex features on the surface of asteroids, such as depressions, canyons, or the necks of contact binary asteroids; therefore direct observations (such as through stellar occultation) are pivotal to understand and characterize non-convex morphological features.

\subsection{Photometry}
\label{sec:photometry}



\begin{figure}
    \includegraphics[width=0.48\textwidth]{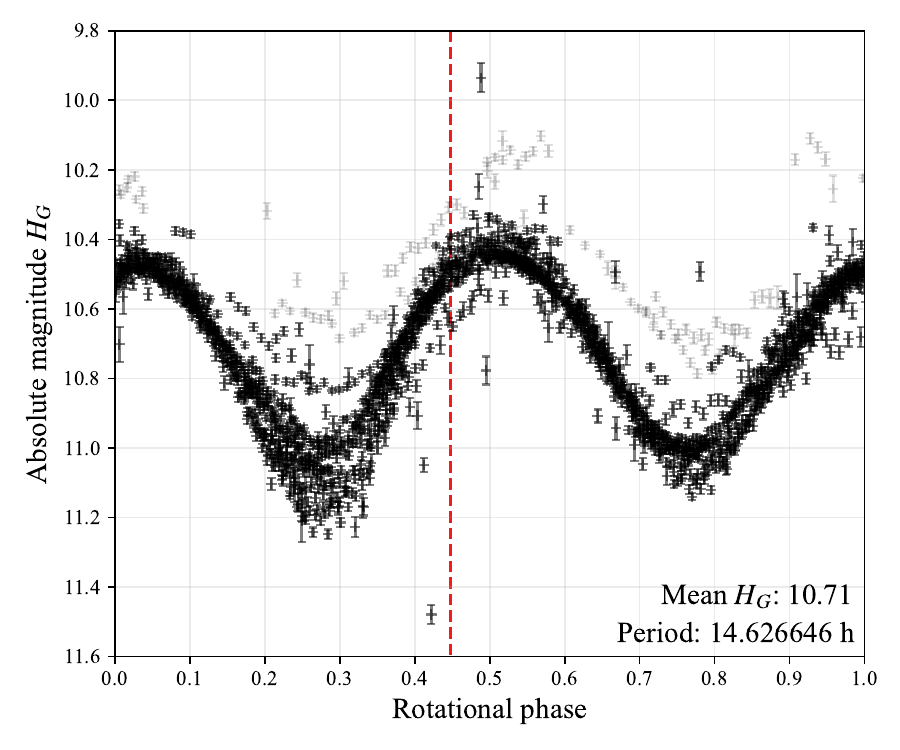}
    \caption{Rotation curve of Chimaera in the \textit{Gaia} G band. Dashed red line shows the rotation phase at the time of the 20251021 occultation.  Certain short periods of suspiciously consistent high magnitude (see §\ref{sec:photometry}) are plotted faintly.}
    \label{fig:photometry}
\end{figure}

Robotic photometry for Chimaera has been collected since September 2025; we use these observations to anchor the rotation phase for the 20251021 occultation and derive the \textit{Gaia} G band albedo for the asteroid. Observations were conducted at the Sierra Remote Observatories with two twin PlaneWave 61.0 cm telescopes operated by the Southwest Research Institute, using a QHY600PH camera with an IMX455 detector. 1903 integrations of 30\,s duration were taken between 20250923 and 20260418. The data were reduced with standard dark and flat subtractions; bias was estimated from overscan and un-illuminated pixels. The calibrated image is binned $3\times3$ before conducting photometry from an aperture with radius equivalent to the average full-width half maximum (FWHM) of each star in the frame. The photometric G band zero-point is derived from the average G magnitude of each star in the frame and applied to the photometry of the asteroid. Binning observations by five-day epochs identify certain periods which systematically possess higher flux but follow the rotation phase of the asteroid well. These are a small subset of the observations and we consider these periods when folding the lightcurve. A Fourier fit dispersion minimization function finds a rotational period of 14.626646 h for the lightcurve; this period may be compared with the values in the JPL Small Body Database: 14.635 h and DAMIT: 14.6241 h. The lightcurve provides a mean G band absolute magnitude $H_G=10.71$ and an absolute magnitude at the time of the 20251021 occultation $H_G\simeq10.5$. The rotation phase-corrected, folded lightcurve can be found in Figure \ref{fig:photometry}, where the suspect regions may be seen as well.

\begin{table}[htbp]
    \centering
    \scriptsize
    \begin{tabular}{c|c|c|c|c|c}
        \hline
        \hline
        Date & Time & Star & $m_*$ & $m_C$ & Chords \\
        $\text{[UTC]}$ & [UTC] & & [mag] & [mag] & \\
        \hline
        20080411 & 07:22 & TYC 7292-01844-1 & 11.49 & 14.76 & 1 \\
        20160426 & 10:45 & UCAC4 243-100271 & 11.85 & 15.11 & 1 \\
        20190222 & 16:02 & TYC 0239-01406-1 & 9.69 & 13.70 & 3 \\
        20240524 & 01:02 & UCAC4 262-174781 & 12.40 & 15.05 & 4 \\
        20241020 & 08:06 & UCAC4 320-181794 & 15.92 & 16.23 & 1 \\
        20251021 & 14:40 & UCAC4 626-004193 & 13.86 & 13.80 & 16 \\
        \hline
        \hline
    \end{tabular}
    \caption{Information for the complete history of occultation observations of (623) Chimaera across the full seventeen-year epoch. The time given is the mid-time for the shadow crossing the surface of the Earth. Apparent V band magnitudes for the star ($m_*$) and Chimaera $m_C$ are given for the time of each occultation.}
    \label{tab:prior_occ_params}
\end{table}

\section{Discussion}
\label{sec:discussion}

\subsection{Morphology}
\label{sec:morphology}

We use the \texttt{SORA} software \citep[Stellar Occultation Reduction and Analysis]{2022MNRAS.511.1167G} to construct a sky-plane plot of the two-dimensional projection of the asteroid's shadow on the fundamental plane. The software accounts for the observers' longitude, latitude, and altitude on Earth, along with the varying times at which the occultation was observed. The resultant sky-plane plot demonstrates the visibility of the target star at each given site along the shadow path. The black lines demonstrate the presence of the star, and the absence of a black line is indicative of the presence of the asteroid. The observed disappearance and reappearance points of the star are traced by a LOWESS function and the interior of the asteroid limb (see §\ref{sec:limb}). Figure \ref{fig:skyplaneplot} can be used to interpret the morphology of Chimaera from these chords. Chords which were negative (i.e. did not observe an occultation) are plotted as straight black lines and provide invaluable constraints on the upper and lower limits of the asteroid size. The gap in coverage in the southern half of the asteroid (due to the absence of two neighboring chords) is apparent, but there is even coverage at roughly 4.4 km intervals throughout the remainder of the profile, with planned clustering at the predicted northern and southern limbs. 

In Figure \ref{fig:ellipses}, we present the known history of occultations by Chimaera. For clarity, we label each chord incrementally to avoid repeating labels and to uniquely identify each chord. For example, the 20240524 event possesses four chords with IDs 20, 21, 22, and 23.

Generally, the projection of Chimaera for the 20251021 event is described well as an ellipse, with a rounder southern limb and a slightly more elongated northern limb. It is very clear from the multi-chord observations of Chimaera that the body is particularly oblate when viewed from certain aspects, in agreement with three-dimensional DAMIT modelling. There are evident deviations from an ellipse which are associated with topography on the surface of Chimaera; these are treated specially in Section \ref{sec:ellipse}.

The presence of a large asteroid family associated with Chimaera suggests a violent collision in prehistory; inter-asteroid collisions have been frequently documented to leave topographical features on other minor planets \citep[etc.]{2012Sci...336..684R, 2024Natur.629.1015L, 2025epsc.conf..408L}, and we suggest that irregularities in the topography of Chimaera as seen through stellar occultation may be a result of the collision which gave rise to the Chimaerid family. One such feature can be seen to the northeast of the 20251021 occultation profile, where the chords 5 and 6 are inset into the asteroid and indicate the presence of a large depression between chords 4 and 7. 

There is another interesting topographical feature in the 20240524 occultation's sky-plane plot which could also be tracing this depression. Chord number 21 appears similar in scale and position angle to the feature traced by chord 11, but, as explored in Section \ref{sec:longitude}, the 20251021 and 20240524 events are separated by an estimated $\sim140\degree$ of projected sub-Earth longitude, indicating that these two features are on nearly opposite hemispheres of the asteroid. This dip detected in the 20240524 event is the only other feature which is distinctly topographical in any of the archival datasets, although there may be some evidence that the 20080411 chord is tracing topography. The western limb for this occultation is significantly closer to the predicted center of the occultation, but there are no nearby chords to compare with to test this hypothesis. There is no strong evidence that this depression is associated with a feature of a contact binary.

Of similar interest is the clustering of chords on the northern and southern limbs of the asteroid, which trace smoothly the tapering of the asteroid's profile before reaching the constraining negatives marked by chords 1 and 18. The clustered chords 2, 3, and four are tracing resolution on the cross- and down-tracks on the $1-2$ kilometer scale, and the especially nearby chords 5 and 6 trace the limb on the sub-kilometer scale.

There is no evidence to suggest the presence of a satellite companion for Chimaera in any of the occultation datasets; either in the lightcurves in the form of a separate occultation event, or detected via a significant deviation from the ephemeris-calculated center of mass for the Chimaera system.

\subsection{Ellipse Fitting}
\label{sec:ellipse}

The morphology of Chimaera, as with many larger asteroids, lends itself well towards a fit with an elliptical shape. Using the \texttt{SORA} analysis software, we perform an $\chi^2$ test of $N=10000$ ellipsoids to each sky-plane plot for Chimaera, which seeks to minimize $\chi^2$ through the relation provided in \citet{2022MNRAS.511.1167G}:

\begin{equation}
    \chi^2=\sum_{i=1}^n\frac{(r_i-r_{ellipse})^2}{\sigma^2_i+\sigma^2_{model}}
\end{equation}

for each datapoint, in which $r_i$ and $r_{ellipse}$ are the positions of the observed chord and the nearest position along the projected test ellipse, respectively. $\sigma_{i}$ is the uncertainty of the datapoint being fit. The value $\sigma_{model}$ is an estimate of the deviation from an ellipse due to topography on the body. From the 20251021 and 20240524 occultations, we can see clear deviations from an ellipse, and estimate the extent of this deviation to be $\sim2$ km; $\sigma_{model}$ is therefore set to 2 km for all occultations. The ellipses are given guess for initial parameters (semi-major axis, oblateness, and position angle) as estimated from the DAMIT model at each predicted occultation mid-time. For the occultations in which only a single chord was observed, we allow for uncertainties of $\pm$ 3.0 km for the semi-major axis, $\pm$ 0.05 for the oblateness, and $\pm$ $30.0\degree$ for the position angle. These relatively tight constraints are necessary to constrain realistic fits given the odd orientation of Chimaera during these events, and the lack of negative constraints for the archival observations; we justify them by our observations of the 20251021 event and comparison with DAMIT modeling at each event time. Even with these constraints the uncertainties for total area for the single-chord events are quite high (approaching or equal to the dispersion of uncertainties allowed; this occurred regardless of the dispersion permitted and we elected to select the aforementioned constraints in light of this to limit results to realistic albeit uncertain values---there are infinite ellipses which may fit a single chord) and we do not attribute as much significance to them as we do to the events on 20190222, 20240524, and 20251021. However, the allowed uncertainties are large enough to consider fits which deviate from the prediction from DAMIT modeling.

\begin{figure*}
    \centering
    \includegraphics[width=0.49\textwidth]{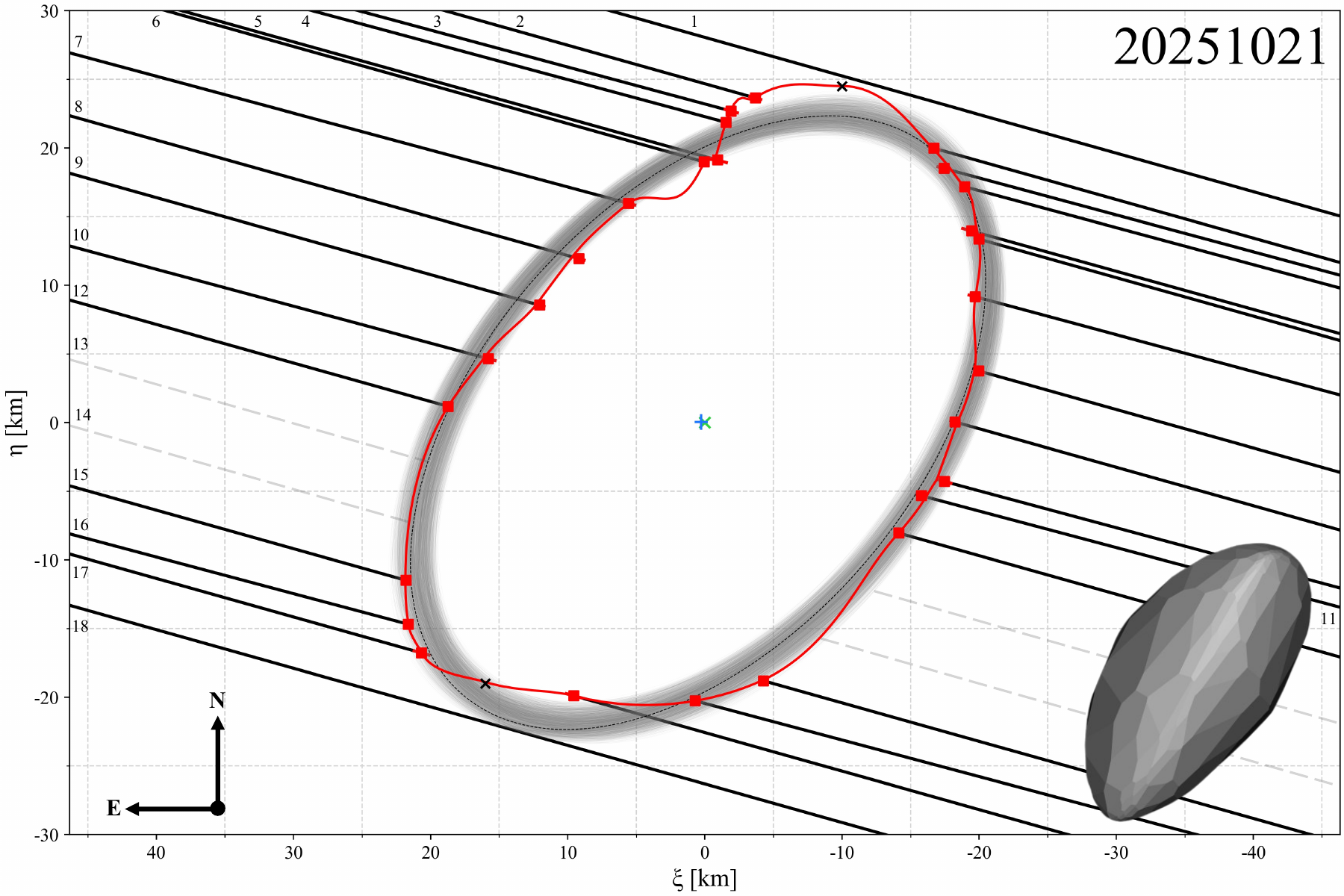}
    \includegraphics[width=0.49\textwidth]{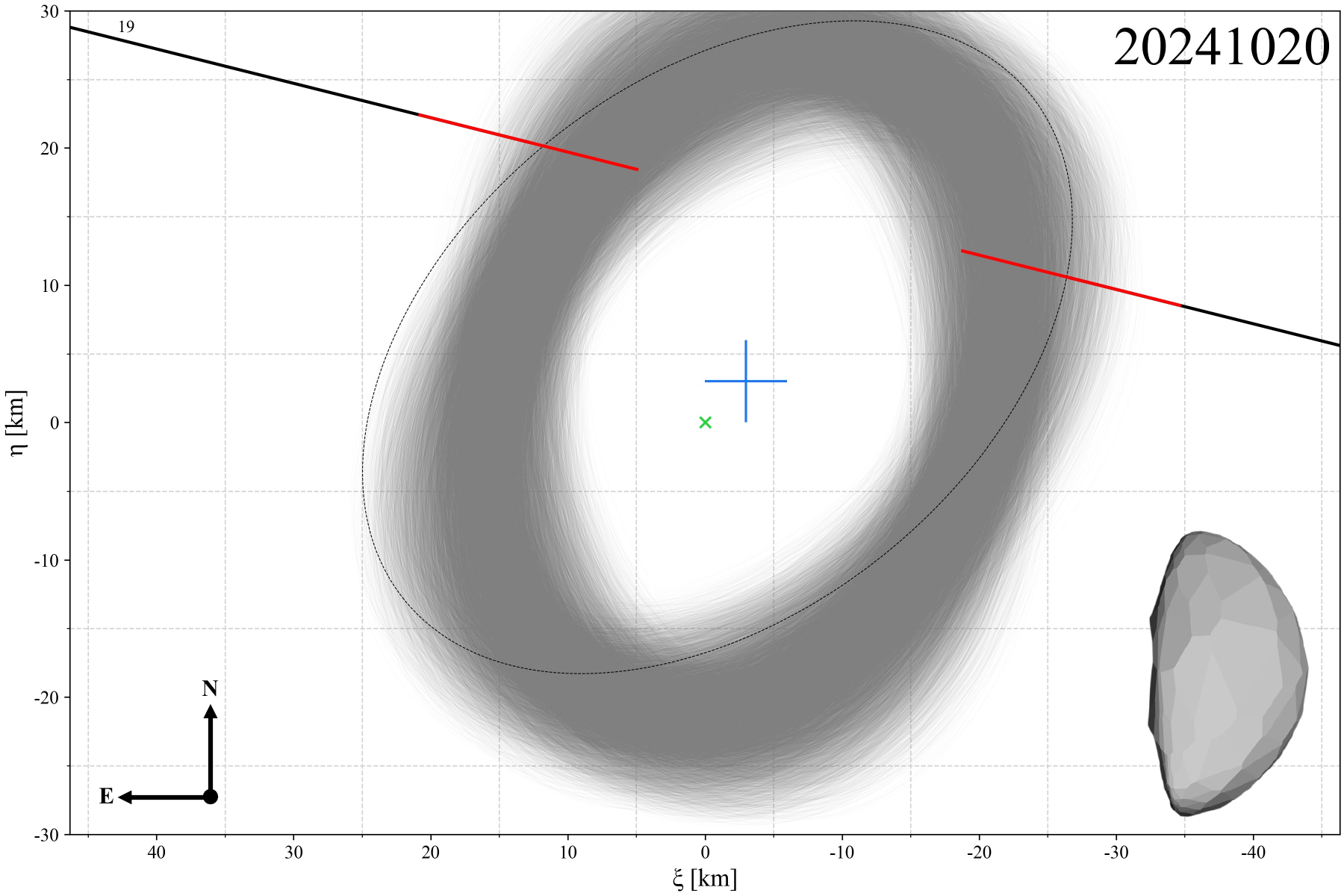}
    \includegraphics[width=0.49\textwidth]{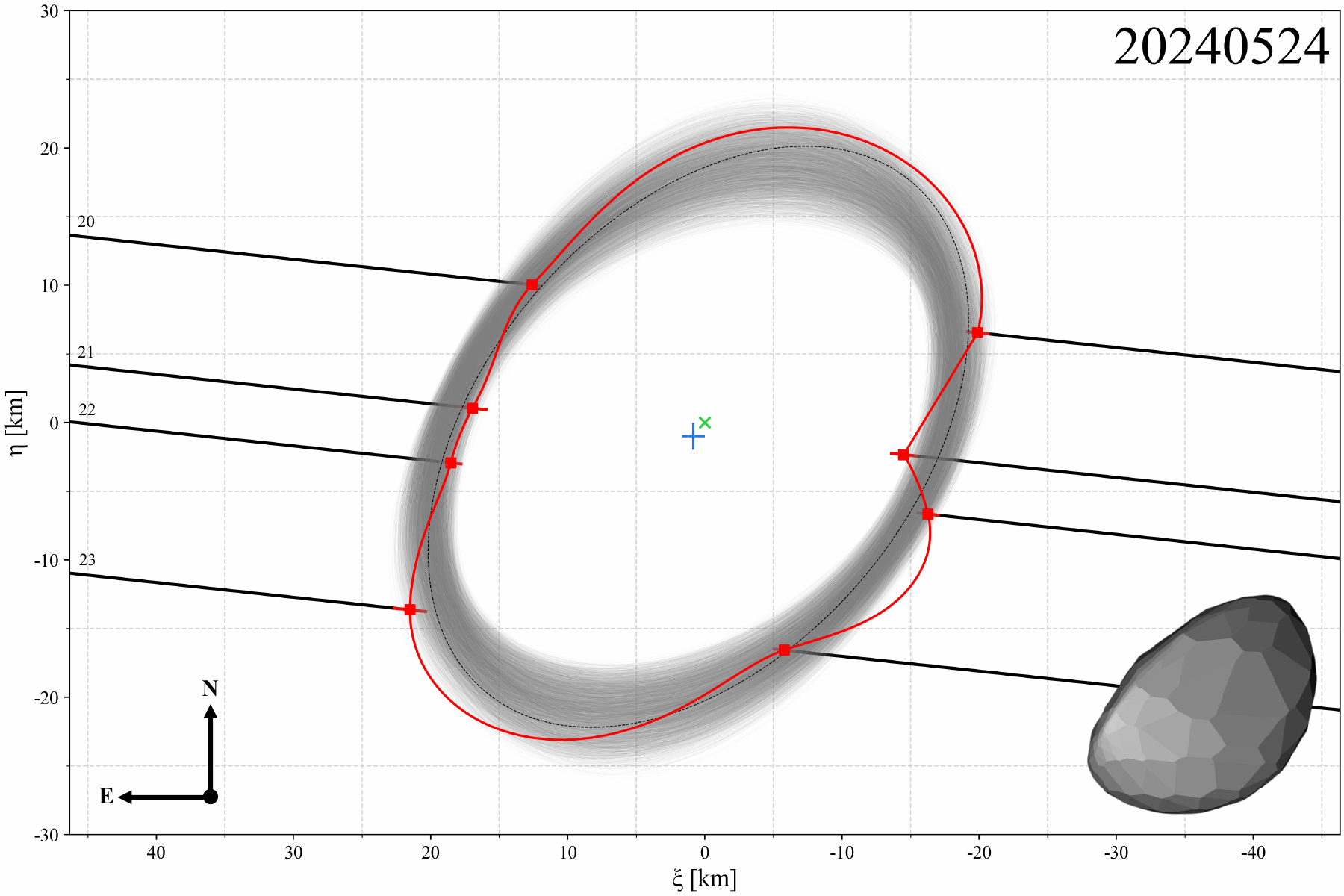}
    \includegraphics[width=0.49\textwidth]{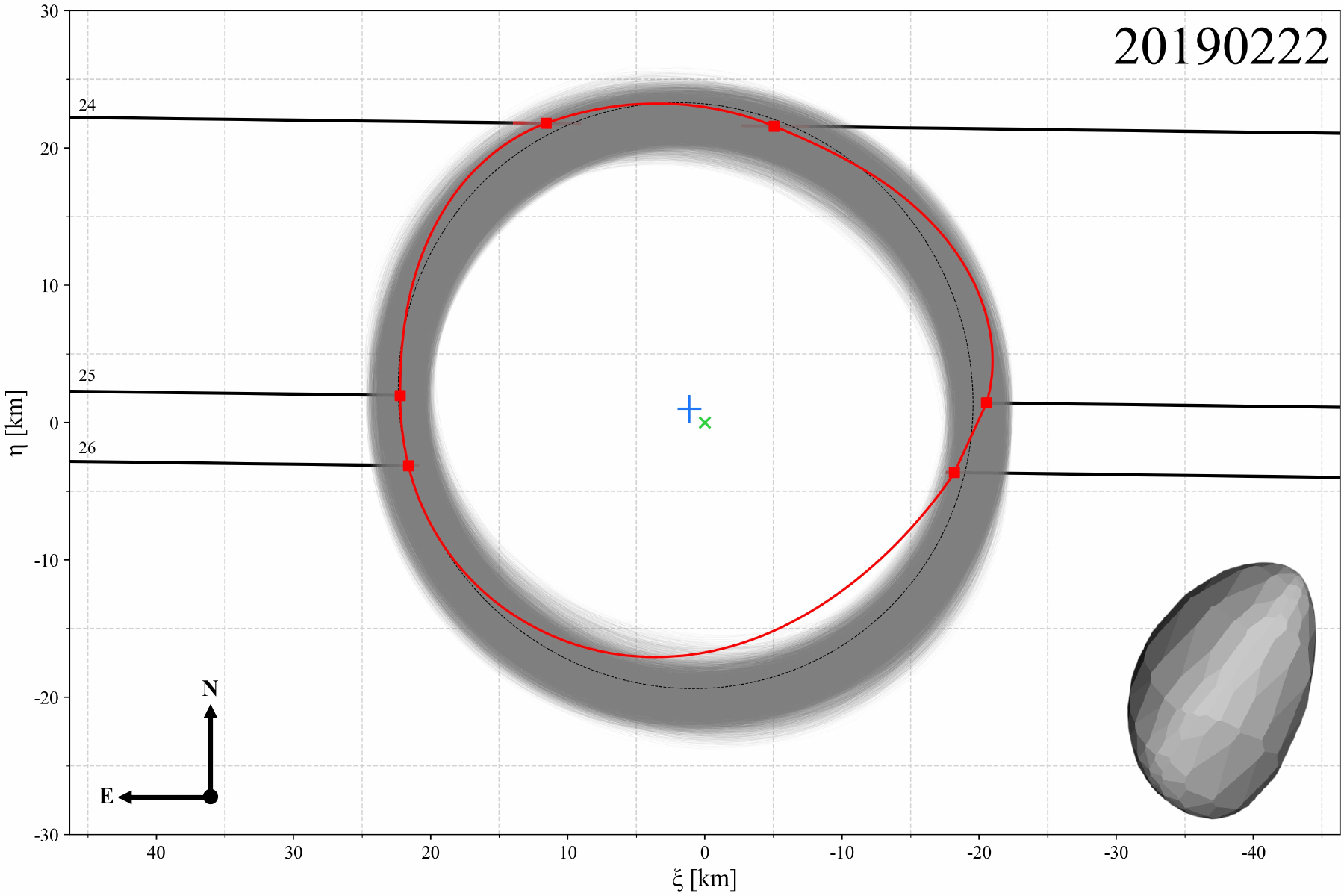}
    \includegraphics[width=0.49\textwidth]{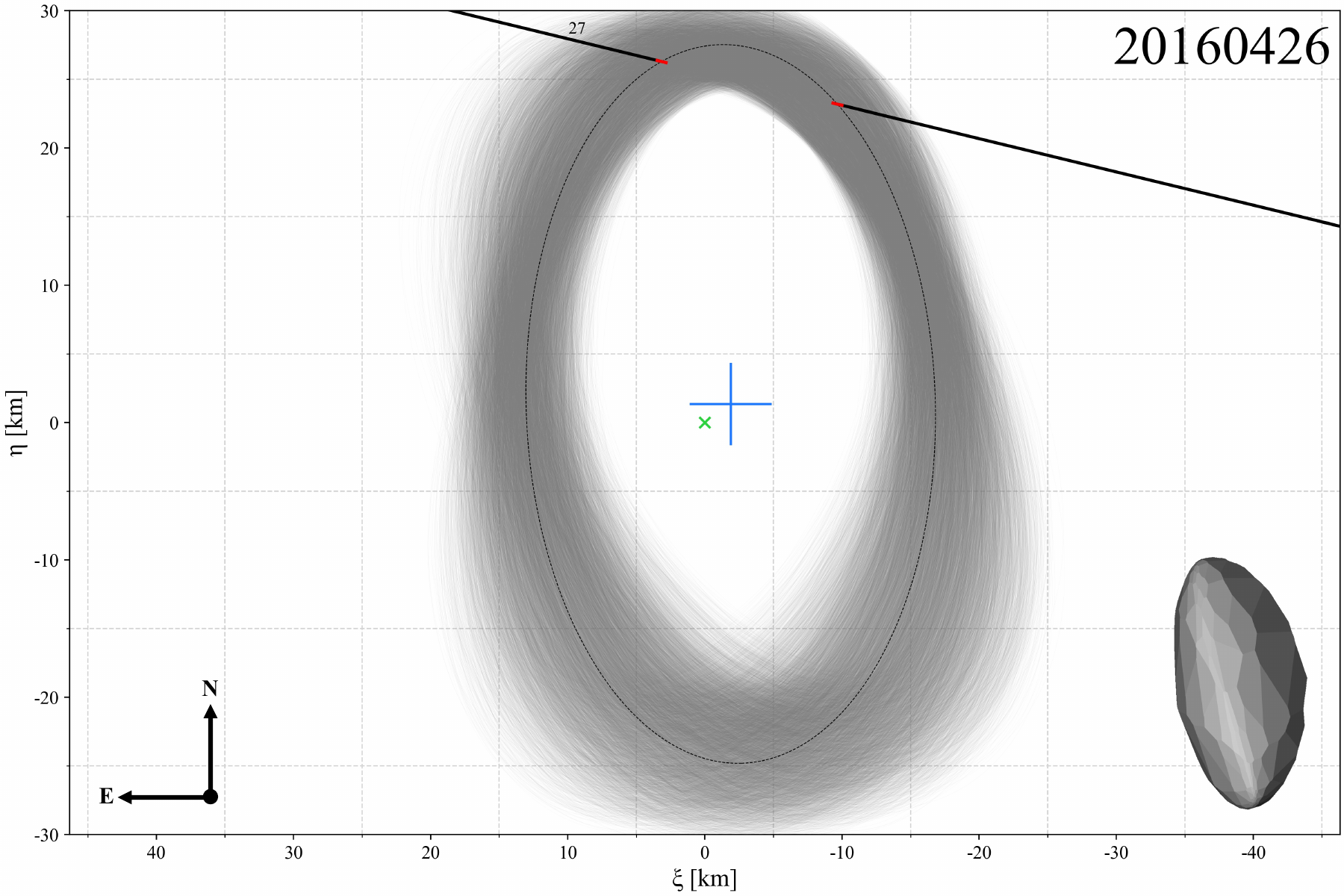}
    \includegraphics[width=0.49\textwidth]{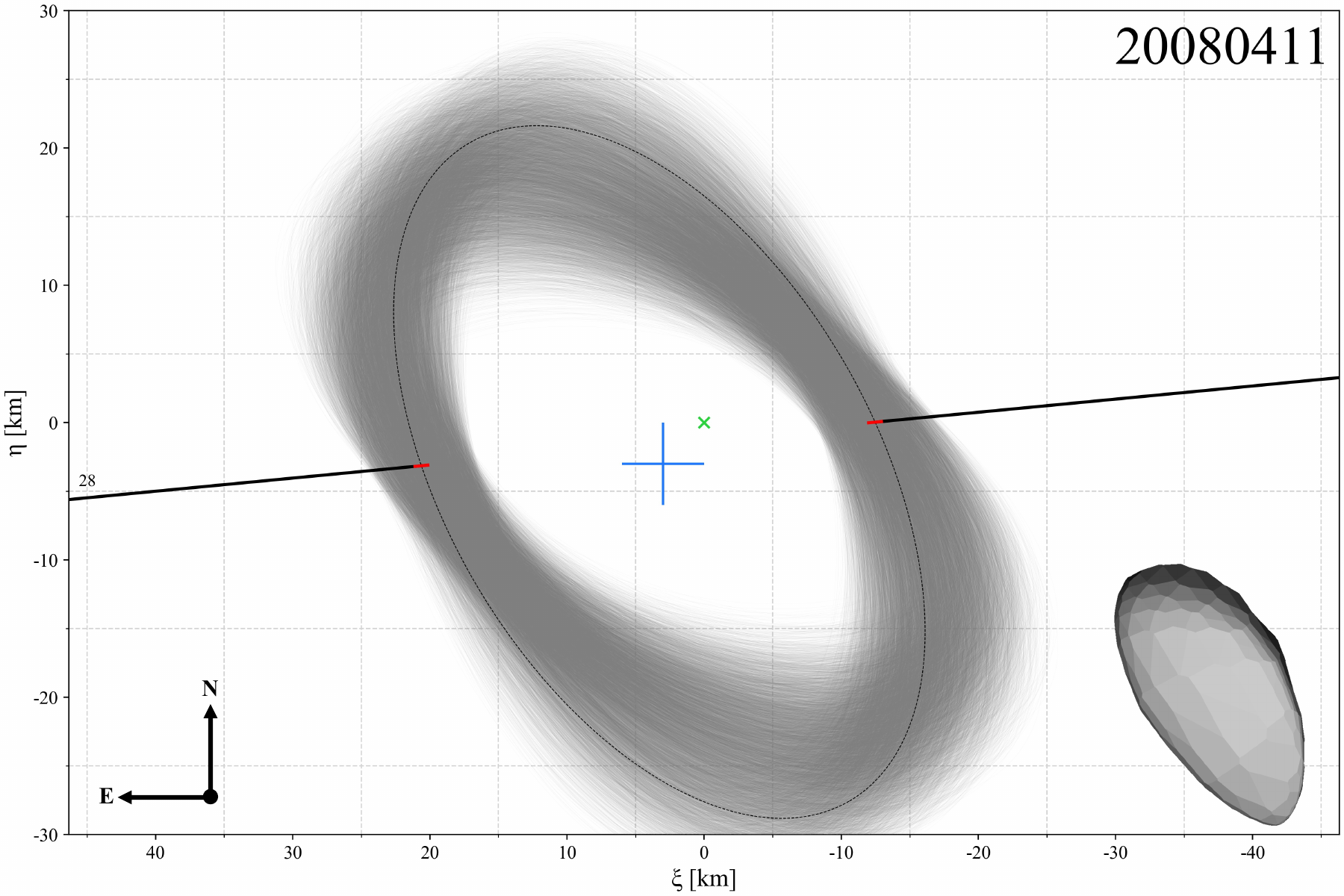}
    \caption{Shape of best-fit ellipse (black dashed line) overlayed on the sky-plane plot for all six observed Chimaera occultations after a run on $N=10000$ tests. All ellipses with $\chi^2$ within 2-$\sigma$ of the best fit are shown in dark gray. The predicted centers from the JPL\#108 ephemeris are marked with a green ``x'' and the average center for the ellipses (and associated uncertainty) are marked with a blue ``+''. For multi-chord observations, the LOWESS-calculated limb fit is plotted in red. The DAMIT model for Chimaera at the time of each occultation is shown in the bottom right of each panel (not to scale).}
    \label{fig:ellipses}
\end{figure*}

The fitted ellipses can be seen superimposed on their respective sky-plane plots in Figure \ref{fig:ellipses}. For the 20251021 occultation, any elliptical fit for which a negative chord intersects is discarded as physically invalid. The equivalent radius for a circle with the same area as our ellipses is calculated by: $R_{eq}=\sqrt{a\cdot b}=20.5\pm0.9$ (equivalent diameter $D_{eq}=40.9\pm1.7$).

\begin{table}[htbp]
    \centering
    \footnotesize
    \begin{tabular}{c|c|c|c|c}
        \hline
        \hline
                       & \multicolumn{3}{c|}{Ellipse fit}                 & Limb fit \\
        \hline
        Date           & \textit{a}   & \textit{b}        & Area         & Area     \\
        $[\text{UTC}]$ & [km]         & [km]              & [km$^2$]     & [km$^2$] \\
        \hline
        20251021       & $26.1\pm0.7$ & $16.0\pm1.0$      & $1316\pm102$  & 1312     \\
        20241020       & $26.5\pm3.0$ & $17.2\pm2.4$      & $1434\pm343$ & \dots    \\
        20240524       & $23.9\pm1.4$ & $16.2\pm1.1$      & $1213\pm110$ & 1296     \\
        20190222       & $22.3\pm2.1$ & $20.0\pm3.0$      & $1401\pm305$ & 1350     \\
        20160426       & $26.2\pm2.9$ & $14.9\pm2.2$      & $1227\pm300$ & \dots    \\
        20080411       & $26.5\pm3.0$ & $13.8\pm2.0$      & $1147\pm282$ & \dots    \\
        \hline
        \hline
    \end{tabular}
    \caption{Output parameters for the fitting performed on each occultation. Uncertainties are arbitrary and subjective for the limb-fits and therefore these area values are provided simply for comparison for the ellipse-fit areas.}
    \label{tab:ellipse_params}
\end{table}

\subsection{Limb Fitting}
\label{sec:limb}

Multi-chord observations of asteroids through stellar occultations can powerfully demonstrate small-scale ($\sim1$ km or less) topography. Therefore a method is desired to accurately map the silhouette, or limb, of the asteroid at the time of the occultation. We follow methodology described in \citep{2025PSJ.....6...61B} to trace the limb of the asteroid as realistically as possible. To begin, the occultation ingress and egress positions with associated uncertainties in the fundamental plane are converted from Cartesian to polar coordinates. A locally-weighted scatterplot smoother (LOWESS or LOESS, \citealt{1986ApOpt..25..203C}) is applied to these data as a method to follow the general trend of the limb, without necessitating that the curve pass through every datapoint when unnatural. We use the \texttt{loess} function in the publically available \texttt{scikit-misc} package in \texttt{Python} to fit a quadratic curve to each datapoint along with only its most close neighbors. The number of neighbors used for the fit is subjectively determined for each multi-chord observation in accordance with the quantity of available neighboring datapoints, but the 20251021 occultation was fit to neighbors within $\pm17.5\degree$ of each datapoint (20240524 and 20190222 were $\pm60\degree$ and $\pm80\degree$, respectively). The LOWESS-computed curve is then converted back into Cartesian coordinates and plotted alongside the true datapoints. The resulting curve is continuous, smooth, and follows the most likely limb of Chimaera, agreeing with the relative uncertainty of each datapoint. To maintain a sensible, realistic solution, two control points were added with large uncertainties. These are inserted to force the LOWESS shape computation to agree with negative chords.

The limb as calculated for the 20251021 occultation is seen in black in Figure \ref{fig:skyplaneplot}. The limb fits conducted for the 20251021, 20240524, and 20190222 occultations are in good agreement with the projected area from ellipse fitting, as can be seen as red contours in Figure \ref{fig:ellipses}. The projected area enclosed by the limb fit for the 20251021 occultation is 1312 km$^2$, which matches well within $1\%$ of the 1310 km$^2$ calculated from ellipse-fitting. The projected areas from fitting the 20240524 and 2019022 occultations are in similar agreement and can be seen in Table \ref{tab:ellipse_params}.

\subsection{Longitudinal Coverage}
\label{sec:longitude}

We refer to the sub-Earth longitude of Chimaera (i.e. which face of Chimaera is facing Earth) at the time of each occultation as presented by DAMIT to further contextualize the occultation history of Chimaera. We attempted to correct for systematic uncertainties in the DAMIT model by comparing the offset in predicted and observed extrema from photometric lightcurves collected contemporaneously near the occultation, but due to the sparsity of photometric data for Chimaera throughout the epoch of occultation data, this was only possible for the 20251021 occultation. Given the success of the DAMIT model in predicting the orientation of Chimaera at each occultation date, we tentatively assume the suggested sub-Earth longitude provided by the model. We incorporate an $\pm10\%$ uncertainty on each area's corresponding longitude.





\begin{figure}
    \centering
    \includegraphics[width=0.48\textwidth]{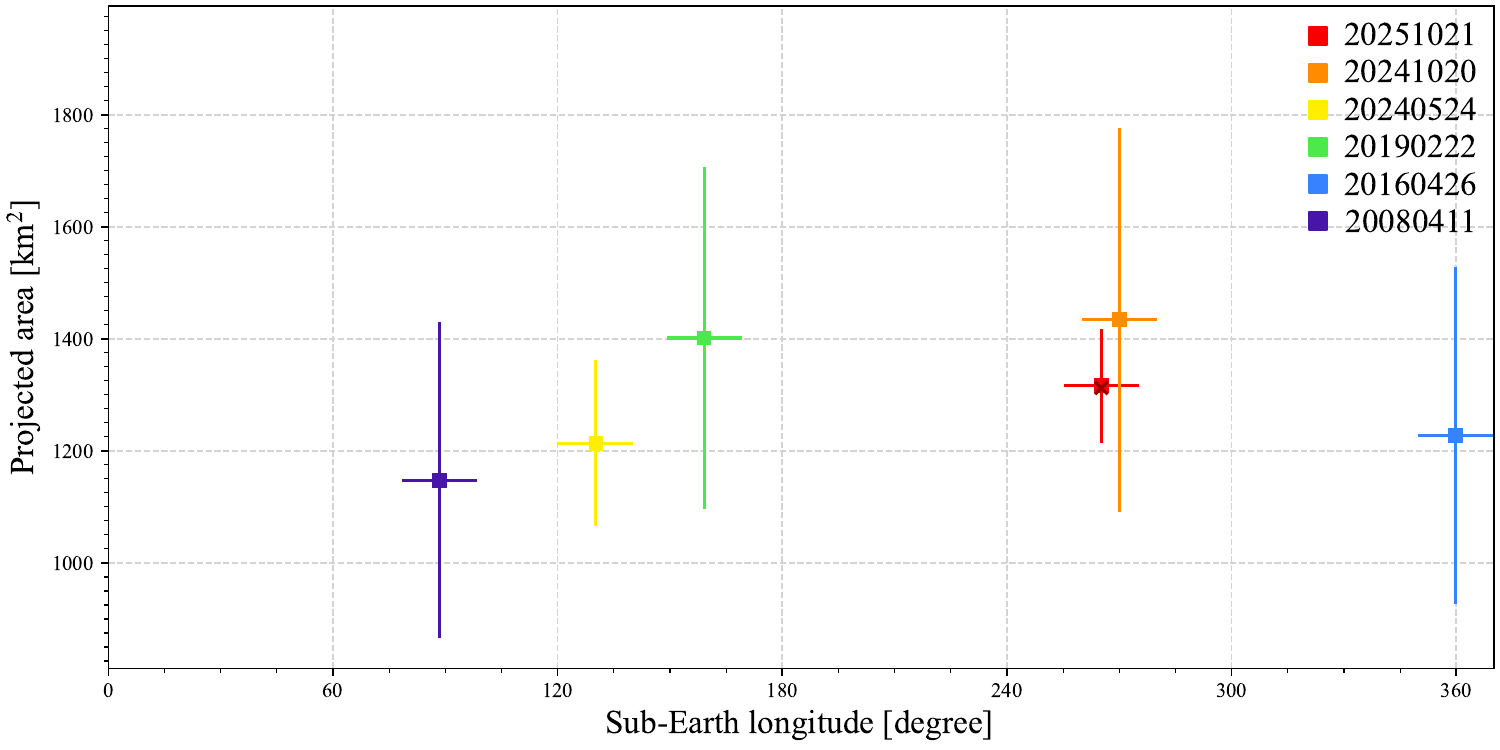}
    \caption{Projected area from each occultation  compared to the sub-Earth longitude of Chimaera at the predicted mid-time for each event. Ellipse fits are marked with squares and error bars, limb fits are marked with an `x'.}
    \label{fig:longitude}
\end{figure}

\subsubsection{Geometric Albedo}
\label{sec:albedo}

With a good constraint on the projected area for Chimaera, we are able to calculate the V-band geometric albedo of Chimaera via the following:

\begin{equation}
    p_X=(\frac{D_{eq}}{K_X}10^{H_{X}/5})^{-2} \\
\end{equation}

with $H_X$ as the absolute magnitude in band $X$, and the constant $K_X$ km derived from the calculation of absolute magnitude for Solar System bodies, itself dependent on the apparent magnitude of the Sun in band $X$ \citep{2007Icar..190..250P, 2021PSJ.....2...32M}. In the V band $K_V=1329\pm10$, and in the \textit{Gaia} G band $K_G=1252\pm10$.

We first test the equivalent diameter $D_{eq}$ of Chimaera as $40.9\pm1.6$ km from the ellipse of best fit for the 20251021 occultation. The absolute magnitude $H_{V}$ is given as 10.83 from IRAS observations, although this is the value averaged across all rotational states and we should account for the absolute magnitude of Chimaera for the rotation phase at the time of the 20251021 occultation. Using our contemporaneous photometry of Chimaera, we calculate that at 07:20 on 20251021 Chimaera was at a rotation phase of $0.93$, which coincides with $H_G\simeq10.5$, and use this value to calculate the geometric albedo $p_G=0.059\pm0.005$. Using the limb-fit area of 1312 km ($D_{eq}=40.9$ km), we arrive at the same conclusion. 

However, given the dependence of the projected area and absolute magnitude on the orientation of Chimaera, it may be more prudent to consider the average projected area across all six occultations. The area-averaged methodology, using equivalent diameter $D_{eq}=40.1\pm1.7$, average absolute G magnitude across all rotation phases $H_G=10.71$, IRAS-derived mean magnitude $H_V=10.83$, gives $p_G=0.051\pm0.004$ and $p_V=0.051\pm0.004$. 
The area-averaged $p_G$ is slightly lower than the $p_G$ calculated for the 20251021 occultation. This may be explained by the significant lack of negative observations for any of the archival observations, and the projected areas from ellipse fitting may be prejudiced towards higher values, and therefore lower albedoes.
These calculations may be compared to a $p_V$ as of $0.0372\pm0.002$ obtained exclusively with IRAS observations \citep{2004PDSS...12.....T}. 

\section{Conclusions}
\label{sec:conclusions}

We have conducted a large campaign to observe an occultation by the main-belt asteroid (623) Chimaera in an effort to probe the asteroid's shape, size, and environment. These measurements will prove valuable to the EMA team as they prepare for flyby in the 2030s. The following conclusions are reached from our observations and our analysis of publicly-available archival data:

Chimaera is an elongated asteroid, and its two-dimensional projection is well-described by an ellipsoidal fit, neglecting kilometer-scale topography. The size of the ellipse fit to the 20251021 event is $1316\pm102$ km$^2$, with an equivalent radius $R_{eq}=20.5\pm0.8$ km. A smooth limb fit encloses an area of $1312$ km$^2$. Averaging the projected areas from geographically unique archival occultations, which cover all hemispheres of Chimaera, suggests an area of $1295\pm75$ km$^2$ with $R_{eq}=20.0\pm1.7$. We find that the three-dimensional DAMIT model of Chimaera is an appropriate representation of the asteroid and can reliably predict the orientation and general shape of the asteroid.
    
Topography is detected on the surface of Chimaera on the kilometer scale in both the 20251021 and 20240524 occultations, potentially tracing the same depression at different rotation phases. There is no evidence for a satellite of Chimaera at this time. We provide a dynamical limb for the asteroid for three independent occultations.

The geometric albedo for Chimaera is confirmed to be low. From the 20251021 occultation we derive a G band albedo of $p_G=0.059\pm0.005$ observation and from all six observations we find $p_G=0.051\pm0.004$ and $p_V=0.051\pm0.004$. This $p_V$ is higher than that determined from IRAS observations.

The results from this campaign and the interpretation of archival occultation results constrain Chimaera's size, shape, and topography, which contextualizes our understanding of the dynamical history of this asteroid.

\begin{acknowledgements}
    
We gratefully acknowledge the support from volunteer observers in the collection of this dataset, particularly Tom Heisey, along with equipment support from the Dark Skies, Bright Kids organization. Financial support from the Parent's Program, Jefferson Trust, and the Student Council at the University of Virginia has been critical in the success of this expedition and we gratefully acknowledge their support of our mission. We thank the following observers for their contribution in compiling the sixteen year epoch of occultation observations prior to this work: Steve Kerr, Graeme McKay, Derek Robson, Luke Broom-Lynne, William Diffin, Hayato Watanabe, Hidehito Yamamura, Akira Asai, Masumi Takimoto, and William Hanna; we extend our thanks to Dave Herald and IOTA for the development and maintenance of the \texttt{Occult4} software for occultation predictions and archival observations. We are grateful for the freely available \texttt{SORA} software developed and maintained by the ERC LuckyStar collaboration. Generative artificial intelligence was not used to write this manuscript.

\end{acknowledgements}

\appendix

\setcounter{table}{0}
\renewcommand{\thetable}{A\arabic{table}}

\begin{table*}[htbp]
    \centering
    \footnotesize
    \begin{tabular*}{0.96\textwidth}{c|c|c|c|c|c|c|c|c|c}
        \hline
        \hline
        ID & Result                    & Latitude    & Longitude    & Altitude & Exposure & Disappearance & Reappearance  & $\sigma_{D\&R}$ & Duration \\
           &                           & [dd.dddddd] & [ddd.dddddd] & [m]      & [s]      & [UTC]         & [UTC]         & [s]             & [s]      \\
        \hline
        1  & \textcolor{red}{NEG}      & 35.351222   & -078.911917  & 67       & 0.300    & \dots         & \dots         & \dots           & \dots    \\
        2  & \textcolor{posgreen}{POS} & 35.314500   & -078.854222  & 56       & 0.300    & 07:15:53.1327 & 07:15:54.5415 & 0.043           & 1.409    \\
        3  & \textcolor{posgreen}{POS} & 35.578972   & -106.125944  & 1885     & 0.167    & 07:19:50.2339 & 07:19:51.9004 & 0.048           & 1.667    \\
        4  & \textcolor{posgreen}{POS} & 35.657972   & -105.232278  & 2019     & 0.070    & 07:19:41.7658 & 07:19:43.6342 & 0.020           & 1.868    \\
        5  & \textcolor{posgreen}{POS} & 35.832417   & -084.617861  & 324      & 0.500    & 07:16:40.1321 & 07:16:42.1323 & 0.069           & 2.000    \\
        6  & \textcolor{posgreen}{POS} & 35.374028   & -079.778333  & 188      & 0.200    & 07:16:00.0701 & 07:16:02.2371 & 0.029           & 2.167    \\
        7  & \textcolor{posgreen}{POS} & 35.446370   & -106.666790  & 1632     & 0.143    & 07:19:54.9131 & 07:19:57.6271 & 0.048           & 2.714    \\
        8  & \textcolor{posgreen}{POS} & 35.341583   & -080.237722  & 160      & 0.200    & 07:16:03.4632 & 07:16:06.6152 & 0.041           & 3.152    \\
        9  & \textcolor{posgreen}{POS} & 35.269611   & -079.924417  & 146      & 0.200    & 07:16:00.9259 & 07:16:04.2037 & 0.035           & 3.278    \\
        10 & \textcolor{posgreen}{POS} & 35.316167   & -106.731056  & 1752     & 0.250    & 07:19:55.5628 & 07:19:59.1298 & 0.051           & 3.623    \\
        11 & \textcolor{blue}{POS-D}   & 34.432306   & -113.242110  & 778      & 0.033    & 07:20:57.6314 & \dots         & 0.040           & \dots    \\
        12 & \textcolor{posgreen}{POS} & 35.158333   & -079.660833  & 160      & 0.200    & 07:15:58.8641 & 07:16:02.4174 & 0.026           & 3.553    \\
        13 & DNO                       & 35.258373   & -106.567415  & 1543     & \dots    & \dots         & \dots         & \dots           & \dots    \\
        14 & DNO                       & 35.011692   & -079.145680  & 79       & \dots    & \dots         & \dots         & \dots           & \dots    \\
        15 & \textcolor{posgreen}{POS} & 35.045806   & -079.713639  & 110      & 0.150    & 07:15:59.8598 & 07:16:02.6701 & 0.030           & 2.810    \\
        16 & \textcolor{posgreen}{POS} & 34.417250   & -112.247750  & 1387     & 0.033    & 07:20:49.6592 & 07:20:51.9009 & 0.012           & 2.242    \\
        17 & \textcolor{posgreen}{POS} & 34.999722   & -079.692222  & 106      & 0.200    & 07:16:00.9828 & 07:16:02.1834 & 0.059           & 1.201    \\
        18 & \textcolor{red}{NEG}      & 34.910306   & -079.212417  & 68       & 0.033    & \dots         & \dots         & \dots           & \dots    \\
        \hline
        \hline
    \end{tabular*}
    \caption{Comprehensive information for each chord, including stellar disappearance and reappearance times with associated uncertainty. The exposure times for each chord is listed.}
    \label{app:d&r}
\end{table*}

\bibliographystyle{aa}
\bibliography{research}

\end{document}